\documentstyle[aps,preprint]{revtex}

\draft

\newcommand{\be}{\begin{eqnarray}}
\newcommand{\ee}{\end{eqnarray}}

\begin{document}
\title{The Markov approximation revisited: inconsistency of the standard quantum Brownian motion model}
\author{$^{1}$Andrea Rocco\footnote{rocco@mailbox.difi.unipi.it} and $^{1,2,3}$Paolo Grigolini\footnote{grigo@soliton.phys.unt.edu}}
\address{$^{1}$Center for Nonlinear Science, University of North
Texas, P.O. Box 305370,
Denton, Texas 76203-305370}
\address{$^{2}$Dipartimento di Fisica dell'Universita' di Pisa, Piazza
Torricelli 2, 56100, Pisa, Italy}
\address{$^{3}$Istituto di Biofisica del Consiglio Nazionale delle
Ricerche,
Via S. Lorenzo 26, 56127 Pisa, Italy}
\date{\today}
\maketitle

\begin{abstract}
We revisit the Markov approximation necessary to derive ordinary Brownian motion from a model widely adopted in literature for this specific purpose. We show that this leads to internal inconsistencies, thereby implying that further search for a more satisfactory model is required.
\end{abstract}

\newpage

\section{Introduction}

According to Penrose \cite{penrose}, the time evolution of a quantum system
should be thought of as the combination of two processes: the $U$-process, {\it i.e}., the unitary time evolution
prescribed by the Schr\"{o}dinger equation, and the $R$-process, namely, the
genuine randomness associated to the collapse of the wave function. This
perspective is very attractive, and many attempts are currently being made
to set the $R$-process on the same dynamical basis as the $U$-process. This purpose is realized by some authors \cite{bl85,ba86,be89} by supplementing the Schr\"{o}dinger equation with stochastic corrections, which are then interpreted as a manifestation of the environment influence.  

As well known \cite{fh}, the path-integral formalism is quite equivalent to
ordinary quantum mechanics. Within this formalism the joint action of the $U$-process and the $R$-process is described by the theory of the path
integral with constraints developed by Mensky \cite{mensky}. We think that a
possible physical meaning of the Mensky approach is made especially
transparent by the interesting recent work of Presilla, Onofrio and Tambini 
\cite{pot,pot2}. These authors prove that the well known method of the
influence functional introduced by Feynmann and Vernon \cite{fv} can be used
to derive the Mensky path integral with constraint as an effect of the
influence of the environment. This is a very interesting result which might
reflect significant elements of truth \cite{tvg95}. The conviction that the
wave-function collapse is generated by an environment and that no such
systems exist in nature as isolated systems is accepted by the majority of
physicists.

In spite of this general consensus, we find some element of inconsistency in
this view. First, we observe that no claim of a genuine derivation of
stochastic properties can be made without involving the Markov
approximation. Adelman \cite{a76} built up a sort of Fokker-Planck equation
equivalent to the generalized Langevin equation and this, in turn, can be
derived from a dynamic picture not involving any Markov approximation \cite
{m65,m65b}, thereby generating the impression that the Markov assumption is
not indispensable to the foundation of a stochastic process. However, as later proved
by Fox \cite{f77}, the non-Markovian Fokker-Planck equation of Adelman {\em %
is not a stochastic process}.

Within a quantum mechanical context the close connection between the Markov
assumption and a stochastic picture is pointed out by the recent work of
Kleinert and Shabanov\cite{ks95}. On the same token we think that a
stochastic picture not implying the Markov approximation is not genuinely
stochastic, in spite of a recent claim to the contrary \cite{ds97}. The
celebrated generalized Langevin equation of Mori \cite{m65,m65b} is an
illuminating example showing that the so called {\em stochastic force} is
actually a deterministic function of the initial conditions and can be
considered as stochastic only if additional assumptions are made, such as
that of coarse graining or of an incomplete information. On
the other hand, these arbitrary assumptions might have essential effects
also on the dynamics: for instance, that of abolishing slow 
tails of the velocity correlation function, conflicting with the exponential
nature of the relaxation of macroscopic variables, a basic tenet of
stochastic physics \cite{vitali}.

This paper is devoted to the discussion of the physical conditions necessary
to establish from within a quantum-mechanical picture a standard
fluctuation-dissipation process for a {\em macroscopic variable}. This
property implies the exponential nature of the relaxation process of the
macroscopic variable, and this is incompatible with both classical \cite{lee}
and quantum mechanics \cite{fonda}. The dynamical approach to the
fluctuation-dissipation process {\cite{vitali} }leads us to conclude that
there are problems even if we disregard the case where the microscopic time
scale is infinite, and the Markov approximation is impossible. In the case
where the Markov approximation seems to be legitimate due to the existence
of a finite time scale, it has the effect of disregarding a weak but
persistent slow tail of the correlation function of the macroscopic
variable. Thus, the Markovian approximation turns out to be equivalent to
the influence of real physical processes, the so called {\em spontaneous
fluctuations} \cite{tvg95}, which are in fact shown to kill these slow tails
\cite{mg98}. The change of perspective is radical, even if it does not
affect the resulting transport equations, and throughout this letter we
shall refer to it as the transition from the {\em subjective} to the {\em 
objective} Markov approximation.

Another way of expressing our aim is as follows. We plan to establish the
intensity of the corrections to ordinary quantum mechanics necessary to make
valid the Markov assumption, which is incompatible with ordinary quantum
mechanics. In literature a scarce attention is devoted to the physical
meaning of the error associated to the Markov approximation. To properly
establish the intensity of this error we interpret the Markov approximation
as a property made genuine by proper corrections to quantum mechanics. Then
we make a balance on the intensity of these corrections. If they turn out to
be too large, we shall consequently judge the error associated with the
Markov approximation to be unacceptable.

To realize this purpose we adopt the Caldeira and Leggett model \cite{cald} and thus our departure point is close to that established by the results of Presilla, Onofrio and Tambini \cite{pot,pot2}, who, as earlier pointed out, showed that the contraction over the irrelevant bath variables yields the Mensky path integral. Similar results have been more recently obtained by Mensky himself, adopting as a bath the internal atomic structure \cite{m87}. To trigger the transition from these subjective Markov assumptions to an objective Markov approximation we adopt the same perspective as that used in \cite{tvg95} to address the problem of the wave-function collapse, through the environment-induced enhancement of spontaneous fluctuations. Furthermore, rather than using the model of Ghirardi, Pearle and Rimini \cite{gpr90} as a source of spontaneous fluctuations, we adopt here a Mensky constraint conceived as a correction to ordinary quantum mechanics.

\section{Markov approximation to friction in the zero-temperature limit}

Notice that the high-temperature assumption of \cite{pot,pot2,cald} with the Mensky constraint as a source of spontaneous fluctuations would make the calculations exceedingly difficult. Thus, we do not address the calculation of fluctuations and we limit ourselves to studying dissipation in the zero-temperature limit. As proved by Caldeira and Leggett \cite{cald}, in the case of harmonic baths the friction term derived from the subjective Markov approximation is independent of temperature. Thus, we make the plausible assumption that even in the presence of the Mensky constraint, the high-temperature friction is the same as that derived by us in the zero-temperature limit (see Section III)\footnote{Of course, after completing the calculation of Section III, yielding a Markov dissipation, we make the implicit assumption that the fluctuations are properly decorrelated by the Mensky constraint so as to recover, in the proper limit, an ordinary fluctuation-dissipation process.}.

Let us consider as in \cite{cald}, an
oscillator of interest interacting with an environment of $N$ oscillators,
with $N\gg 1$. The Lagrangians describing this system are given by the
following expressions: 
\be
&&{\cal L}_0(q) = \frac{1}{2} m_0 \dot{q}^2 - \frac{1}{2} m_0 \omega_0^2 q^2
\\
&&{\cal L}_B(Q_i) = \sum_{j=1}^N \left\{\frac{1}{2} m_B \dot{Q}_j^2 - \frac{1%
}{2} m_B \omega_j^2 Q_j^2 \right\}  \label{lag} \\
&&{\cal L}_I(q,Q_j) = q \sum_{j=1}^N g_j Q_j.
\ee
Thus the resulting quantum mechanical motion of the whole system is
described by the following quantum mechanical propagator: 
\be
&&\langle q_{F},Q_{1}^{F},...,Q_{N}^{F},T|q_{I},Q_{1}^{I},...,Q_{N}^{I},0\rangle =\int[dq]\exp \left\{ \frac{i}{\hbar }\int_{0}^{T}dt\left[ \frac{1}{2}m_{0}\dot{q}^{2}-\frac{1}{2}m_{0}\omega _{0}^{2}q^{2}\right] \right\}  \nonumber \\
&& \qquad \qquad \qquad \qquad \qquad \times \prod_{j=1}^{N}\int [dQ_{j}]\exp \left\{ \frac{i}{\hbar }\int_{0}^{T}dt\left[ \frac{1}{2}m_{B}\dot{Q}_{j}^{2}-\frac{1}{2}m_{B}\omega
_{j}^{2}Q_{j}^{2}+g_{j}qQ_{j}\right] \right\} .  \label{mod}
\ee
Let us use the contracted description already adopted by Feynman and
co-workers \cite{fh,fv}. This means the reduced density matrix $\rho _{A}$ 
\cite{cald} for the system of interest defined by: 
\be
\rho _{A}(q_{F},q_{I},T)=\int dq_{F}^{\prime },dq_{I}^{\prime}K(q_{F},q_{I},T;q_{F}^{\prime },q_{I}^{\prime},0)\rho
_{A}(q_{F}^{\prime },q_{I}^{\prime},0).
\ee
Here $K$ denotes the superpropagator 
\be
K(q_F,q_I,T;q_F^{\prime},q_I^{\prime},0)=\int [dq_{1}][dq_{2}]\exp
\left\{ \frac{i}{\hbar }\big[S_{0}(q_{1})-S_{0}^{*}(q_{2})\big]\right\} 
{\cal F}[q_{1},q_{2}],  \label{superprop}
\ee
where $S_0$ is the action related to ${\cal L}_0$ and 
${\cal F}$ is the influence functional \cite{fh,fv} whose explicit
expression is 
\be
{\cal F}[q_{1},q_{2}]=\exp \left\{ -\frac{1}{\hbar }
\int_{0}^{T}dt[q_{1}(t)-q_{2}(t)]I_{R}(t)-\frac{i}{\hbar }%
\int_{0}^{T}dt[q_{1}(t)-q_{2}(t)]I_{I}(t)\right\},  \label{influe}
\ee
with the two convolution integrals $I_{R}(t)$ and $I_{I}(t)$ given by: 
\be
&&I_R(t) = \int_0^t dt^{\prime} \alpha_R(t-t^{\prime}) [q_1(t^{\prime}) -
q_2(t^{\prime})],  \label{conv1} \\
&&I_I(t) = \int_0^t dt^{\prime} \alpha_I(t-t^{\prime}) [q_1(t^{\prime}) +
q_2(t^{\prime})].  \label{conv2}
\ee
The two memory kernels $\alpha _{R}(t-t^{\prime })$ and $\alpha
_{I}(t-t^{\prime })$ present in (\ref{conv1}) and (\ref{conv2}) in the zero-temperature limit are given by the expressions 
\be
&&\alpha_R(t-t^{\prime}) = \frac{\eta}{\pi} \int_0^{\Omega} d \omega \omega
\cos \omega(t - t^{\prime}),  \label{k21} \\
&&\alpha_I(t-t^{\prime}) = - \frac{\eta}{\pi} \int_0^{\Omega} d \omega
\omega \sin \omega(t - t^{\prime}),  \label{k22}
\ee
where $\eta$ is the friction and $\Omega$ is a cut-off frequency introduced assuming ohmic environment \cite{cald2}.

We are now in a position to discuss the problems raised by the subjective
Markov approximation on the two kernels (\ref{k21}) and (\ref{k22}). First,
we observe that the Markov approximation is expressed by: 
\be
I(t)=\int_{0}^{t}dt^{\prime }f(t^{\prime }){\cal K}(t-t^{\prime
})=\int_{0}^{t}d\tau f(t-\tau ){\cal K}(\tau )\simeq f(t)\int_{0}^{\infty
}d\tau {\cal K}(\tau ).  \label{dec}
\ee
According to the traditional wisdom, the closer ${\cal K}(\tau )$ to $%
\delta(\tau )$, the more accurate this approximation is expected to be. From
within the perspective adopted in this paper, however, we have to establish
a clearcut separation between the case $\Omega =\infty$ and the case $\Omega
< \infty$, regardless of how large $\Omega $ might be in this latter case.
This is so because we know\cite{vitali} that, in accordance with the work of
Ref. \cite{lee,fonda}, the relaxation process of a variable made unstable by
a bath with a bounded spectrum, cannot be exactly exponential. At long
times, after the completion of the exponential decay of the macroscopic
variable, non-exponential tails appear\cite{vitali}. As weak as these tails
are made by increasing the value of $\Omega$, their existence is
incompatible with the claim that a fluctuation-dissipation relation has been derived.

The ideal case of $\Omega =\infty $, on the contrary, would make the
property of (\ref{dec}) exact, and there would be no conflict with the requirements
of stochastic theory. However, we think that the recourse to the property $
\Omega =\infty$ is a sort of subterfuge and that this choice and the
thermodynamical properties of crystals are at odds. 

Let us discuss now in this light the memory kernels of the case under study.
Let us consider first the real part of the memory kernel $\alpha
_{R}(t-t^{\prime })$. This corresponds to fluctuations, and in accordance
with the known low-temperature predictions\cite{west} we expect a strong non
Markovian behavior to emerge out of it. In fact, renaming $\tau =t-t^{\prime
}$, and integrating (\ref{k21}), we get: 
\be
\alpha _{R}(\tau )=\frac{\eta }{\pi }\left[ \Omega \frac{\sin \Omega \tau }{%
\tau }-2\frac{\sin ^{2}(\Omega \tau /2)}{\tau ^{2}}\right] .
\ee
This is integrable, and the adoption of the Markovian approximation, which is
apparently possible, would lead to the puzzling result: 
\be
\int_{0}^{\infty }\alpha _{R}(\tau )=\frac{\eta \Omega }{2}-\frac{\eta
\Omega }{2}=0.
\ee
Actually, in accordance with \cite{west,hpz}, this is a case of infinite
memory strikingly departing from the $\delta $-function condition which is
reached with $\Omega \rightarrow \infty $ by the dissipation part. Not even
the subjective Markov approximation is admitted in this case.

Then let us consider $\alpha _{I}(\tau )$. In this case, on the basis of 
\cite{cald} we expect a Markovian behavior, temperature independent, to be
legitimate at a subjective level. The following calculations confirm our
expectation. From Eq. (\ref{k22}), we get 
\be
\alpha _{I}(\tau )=\eta \frac{d}{d\tau }\left( \frac{1}{\pi }\frac{\sin
\Omega \tau }{\tau }\right).
\ee
By using 
\be
\lim_{\Omega \rightarrow \infty }\frac{1}{\pi }\frac{\sin \Omega \tau }{\tau 
}=\delta (\tau ),
\ee
we get the expression: 
\be
\alpha _{I}(\tau )=\eta \delta ^{\prime }(\tau ).
\ee
Therefore the convolution integral (\ref{conv2}) becomes: 
\be
I_{I}(t)=-\eta \delta (0)[q_{1}(t)+q_{2}(t)]+\frac{\eta }{2}[\dot{q}_{1}(t)+
\dot{q}_{2}(t)].
\ee
The adoption of this approach makes it possible to evidentiate that, as
earlier pointed out, the case $\Omega =\infty $ yields an exact result. In
the case $\Omega < \infty$ the same result can be obtained using the
approximation of (\ref{dec}): 
\be
I_{I}(t) &=&\eta \int_{0}^{t}dt^{\prime }\frac{d}{d(t-t^{\prime })}\left( 
\frac{1}{\pi }\frac{\sin \Omega (t-t^{\prime })}{t-t^{\prime })}\right)
[q_{1}(t^{\prime })+q_{2}(t^{\prime })]  \nonumber \\
&\simeq&-\eta \left. \frac{1}{\pi }\frac{\sin \Omega (t-t^{\prime })}{t-t^{\prime
}}[q_{1}(t^{\prime })+q_{2}(t^{\prime })]\right| _{0}^{t}+\frac{\eta }{\pi }[\dot{q}_{1}(t)+\dot{q}_{2}(t)]\int_{0}^{\infty }d\tau \frac{\sin \Omega \tau 
}{\tau }  \nonumber \\
&=&-\eta \frac{\Omega }{\pi }[q_{1}(t)+q_{2}(t)]+\frac{\eta }{2}[\dot{q}_{1}(t)+\dot{q}_{2}(t)].  \label{deco}
\ee
This result coincides with that resting on the condition $\Omega = \infty$. However, as earlier remarked, the Markov approximation is equivalent to disregarding weak but very slow tails \cite{vitali}.

In both cases, the dissipative part of the influence functional is therefore written as \cite{west,hpz}: 
\be
{\cal F}^{(diss)}[q_{1},q_{2}] &=& \exp \left\{\frac{i}{\hbar }
\int_{0}^{T}dt\eta \delta (0)[q_{1}^{2}(t)-q_{2}^{2}(t)] \right.  \nonumber
\\
&& - \left. \frac{i}{\hbar } \int_{0}^{T}dt\frac{\eta }{2}%
[q_{1}(t)-q_{2}(t)][\dot{q}_{1}(t)+\dot{q}_{2}(t)]\right\}.  \label{tretre}
\ee
Notice that the divergence associated with $\delta (0)$ can be settled 
either by assigning to the Lagrangian a proper counter-term \cite{cald2,grab} or assuming a correlated
initial condition for the total system \cite{pat}.

\section{From the subjective to the objective markov approximation}

We want to explore here the possibility that the problems with the choice of
a finite value of $\Omega $ might be solved assuming that the bath
oscillators are subjected to a Mensky measurement process \cite{mensky,pot}.
Adopting the perspective of Onofrio, Presilla and Tambini \cite{pot,pot2},
these measurement processes should be traced back to the interaction between
each bath oscillator and its own bath. This would not be yet satisfactory
because the finite $\Omega $ problem would now affect the baths of the bath:
This would be the beginning of an endless chain of baths of baths. We
truncate this endless chain by assuming that the measurement process on the
bath oscillator is an expression of a generalized version of quantum
mechanics, an expression of spontaneous fluctuations, in the spirit of the
theory of Ghirardi, Pearle and Rimini \cite{gpr90}. Of course, also the
oscillator of interest should undergo the direct influence of this process.
However, we are also making the assumption that the corrections to ordinary
quantum mechanics are extremely weak, thereby implying that the direct influence of the Mensky
measurement process on the oscillator of interest can be safely neglected.

On these premises we are naturally led to adopt the following quantum
mechanical propagator: 
\be
&&\langle q_{F},Q_{1}^{F},...,Q_{N}^{F},T|q_{I},Q_{1}^{I},...,Q_{N}^{I},0\rangle
_{\Gamma } = \int [dq]\exp \left\{ \frac{i}{\hbar }\int_{0}^{T}dt\left[ 
\frac{1}{2}m_{0}\dot{q}^{2}-\frac{1}{2}m_{0}\omega _{0}^{2}q^{2}\right]
\right\}  \nonumber \\
&& \qquad \times \prod_{j=1}^{N}\int [dQ_{j}]\exp \left\{ \frac{i}{\hbar }%
\int_{0}^{T}dt\left[ \frac{1}{2}m_{B}\dot{Q}_{j}^{2}-\frac{1}{2}m_{B}\omega
_{j}^{2}Q_{j}^{2}+i\hbar k(\omega _{j})(Q_{j}-a_j)^{2}+g_{j}qQ_{j}\right]
\right\},  \label{mod2}
\ee
where 
$a_j = a_j(t)$ is a function which expresses the result of the measurement process taking place on the $j$-th bath oscillator and $k(\omega_j)$ is the strenght of such a process. For the sake of generality, we assume the strenght of this
measurement process to depend on the bath-oscillator frequency. After
several trials we established that the most convenient form to adopt for our
purposes is the following linear dependence: 
\be
k(\omega _{j})=\frac{\Gamma m_{B}}{\hbar }\left( \omega _{j}-\frac{i\Gamma }{%
2}\right).  \label{kappa}
\ee

The main tenet of all theoretical derivations of fluctuation-dissipation
processes \cite{g85} is that the fluctuation-dissipation processes are
perceived at a contracted level of description, and that these processes are
nothing but a manifestation of the interaction with a bath, whose elementary
constituents are not directly observed. This makes tempting the adoption of
Markov approximation, as an expression of this lack of knowledge. However,
in the same way as the information approach to statistical mechanics by
Jaynes \cite{ja} implies probabilistic ingredients which might be foreign to
classical mechanics, the Markov approximation is equivalent to an elementary
randomness, which would be foreign to quantum mechanics, if this is not
supplemented by the action of $R$-processes\cite{penrose}. The purpose of
this letter is to establish the amount of this elementary randomness with
the help of the Mensky formalism.

Note that the assumption that the bath is found in a condition of
equilibrium, expressed by a canonical distribution at temperature $T$ is a
very strong assumption, already implying the inclusion of thermodynamical
arguments. As discussed in Section IV, we have in mind a perspective, based
on chaotic dynamics, where thermodynamics can be really reduced to dynamics,
and this should prevent us from adopting this strong assumption. At the same
time, this assumption, based on the fact that dynamics and statistics are
decoupled the ones from the others, makes it possible to adopt a harmonic
thermal bath, whose regular dynamics has the effects of rendering
excessively large, as we shall see, the corrections to quantum mechanics
necessary to make the Markov assumption possible.

In conclusion, using again the contracted description already adopted in the
previous Section, we get for the influence functional the expression: 
\be
{\cal F}_{\Gamma }[q_{1},q_{2}] &=&\exp \left\{ -\frac{1}{\hbar }
\int_{0}^{\infty }d\omega \frac{g^{2}(\omega )}{2m_{B}\omega }\left( \frac{dN
}{d\omega }\right) \left[ \int_{0}^{T}dt\int_{0}^{t}dt^{\prime
}q_{1}(t)q_{1}(t^{\prime })e^{-i\omega (t-t^{\prime })-\Gamma (t-t^{\prime
})}\right. \right.  \nonumber \\
&&+\left. \left. \int_{0}^{T}dt\int_{0}^{t}dt^{\prime
}q_{2}(t)q_{2}(t^{\prime })e^{i\omega (t-t^{\prime })-\Gamma (t-t^{\prime
})}\right. \right.  \nonumber \\
&&\qquad -\left. \left. \int_{0}^{T}dt\int_{0}^{t}dt^{\prime
}q_{1}(t)q_{2}(t^{\prime })e^{i\omega (t-t^{\prime })-\Gamma (2T-t-t^{\prime
})}\right. \right.  \nonumber \\
&&\qquad\qquad-\left. \left. \int_{0}^{T}dt\int_{0}^{t}dt^{\prime
}q_{2}(t)q_{1}(t^{\prime })e^{-i\omega (t-t^{\prime })-\Gamma
(2T-t-t^{\prime })}\right] \right\} ,  \label{inf}
\ee
where we have neglected terms of order $\Gamma $ or higher related with the
definition (\ref{kappa}). Notice the presence of the two memory kernels
depending on the sum of the two times $t$ and $t^{\prime }$ rather than on
their difference.

Eq. (\ref{inf}) can be rewritten under a form similar to (\ref{influe}): 
\be
{\cal F}_{\Gamma }[q_{1},q_{2}] &=&\exp \left\{ -\frac{1}{\hbar }%
\int_{0}^{T}dt[q_{1}(t)I_{R,1}^{(\Gamma) }(t)+q_{2}(t)I_{R,2}^{(\Gamma)
}(t)-q_{1}(t)J_{R,2}^{(\Gamma) }(t)-q_{2}(t)J_{R,1}^{(\Gamma) }(t)]\right. 
\nonumber \\
&&\left. -\frac{i}{\hbar }\int_{0}^{T}dt[q_{1}(t)I_{I,1}^{(\Gamma)
}(t)-q_{2}(t)I_{I,2}^{(\Gamma) }(t)+q_{1}(t)J_{I,2}^{(\Gamma)
}(t)-q_{2}(t)J_{I,1}^{(\Gamma) }(t)]\right\},  \label{fg}
\ee
where 
\be
I_{\lambda ,k}^{(\Gamma) }(t)=\int_{0}^{t}dt^{\prime }q_{k}(t^{\prime
})\alpha_{\lambda }^{(\Gamma) }(t-t^{\prime })  \label{mela}
\ee
and 
\be
J_{\lambda ,k}^{(\Gamma) }(t)=e^{-2\Gamma (T-t)}I_{\lambda ,k}^{(\Gamma) }(t),
\ee
with $k=1,2$ and $\lambda =R,I$. The modified kernels $\alpha _{R}^{(\Gamma)
}(\tau )$ and $\alpha _{I}^{(\Gamma) }(\tau )$ are defined as follows: 
\be
&&\alpha _{R}^{(\Gamma) }(\tau )=\alpha _{R}(\tau )e^{-\Gamma \tau }=\frac{%
\eta }{\pi }\int_{0}^{\Omega }d\omega \omega \cos \omega \tau e^{-\Gamma
\tau } \\
&&\alpha _{I}^{(\Gamma) }(\tau )=\alpha _{I}(\tau )e^{-\Gamma \tau }=-\frac{%
\eta }{\pi }\int_{0}^{\Omega }d\omega \omega \sin \omega \tau e^{-\Gamma
\tau },
\ee
that is, neglecting terms of order $\Gamma $: 
\be
&&\alpha _{R}^{(\Gamma) }(\tau )=\frac{\eta }{\pi }\left[ \Omega \frac{\sin
\Omega \tau }{\tau }-2\frac{\sin ^{2}(\Omega \tau /2)}{\tau ^{2}}\right]
e^{-\Gamma \tau } \label{exp1} \\
&&\alpha _{I}^{(\Gamma) }(\tau )=\frac{\eta }{\pi }\frac{d}{d\tau }\left( 
\frac{\sin \Omega \tau }{\tau }e^{-\Gamma \tau }\right).  \label{exp2}
\ee
Both memory kernels decay exponentially in time, a fact that according to
the earlier remarks fully legitimates the Markov behavior of the system. Let
us see it in some more detail. As far as $\alpha _{R}^{(\Gamma) }$ is
concerned, its scale becomes now 
\be
\tau _{R}=\int_{0}^{\infty }\alpha _{R}^{(\Gamma) }(\tau )d\tau =\frac{\eta }{%
2\pi }\Gamma \ln \left( \frac{\Gamma ^{2}+\Omega ^{2}}{\Gamma ^{2}}\right) =%
\frac{\eta }{2\pi }{\cal O}(\Gamma ),
\ee
that is, it is different from zero for ${\cal O}(\Gamma )$. In the limit $%
\Gamma \rightarrow 0$, $\tau _{R}\rightarrow 0$, restating the already
discussed non-Markovian properties.

More interesting is the case of $\alpha_I^{(\Gamma)}$. For $\Gamma$ small but
finite, the Markov approximation becomes now rigorous because of the
presence of the $e^{-\Gamma \tau}$ term. We get, for the convolution
integral (\ref{mela}): 
\be
I_{I,k}^{(\Gamma)}(t) & = & \frac{\eta}{\pi} \int_0^t d \tau \frac{d}{d \tau}
\left( \frac{\sin \Omega \tau}{\tau} e^{-\Gamma \tau} \right) q_k(t - \tau) 
\nonumber \\
& = & - \frac{\eta}{\pi} \Omega q_k(t) + \frac{\eta}{\pi} \left(\int_0^{%
\infty}d \tau \frac{\sin \Omega \tau}{\tau} e^{-\Gamma \tau}\right) \dot{q}%
_k(t)  \nonumber \\
& = & - \frac{\eta}{\pi} \Omega q_k(t) + \frac{\eta}{\pi} \arctan\left(\frac{%
\Omega}{\Gamma}\right)\dot{q}_k(t)
\ee
Notice that we could get this result without carrying out the limit $\Omega
\rightarrow \infty$. In the limit $\Gamma \rightarrow 0$, 
\be           
&&I_{I,k}^{(\Gamma)}(t) \rightarrow -\frac{\eta}{\pi} \Omega q_k(t) + \frac{%
\eta}{2} \dot{q}_k(t) \\
&&J_{I,k}^{(\Gamma)}(t) \rightarrow -\frac{\eta}{\pi} \Omega q_k(t) + \frac{%
\eta}{2} \dot{q}_k(t)
\ee       
and from (\ref{fg}), Eq. (\ref{deco}) and (\ref{tretre}) are recovered, with
$\Omega$ replacing $\delta(0)$.

\section{Intensity of the corrections to ordinary quantum mechanics
necessary to the objective Markov approximation}

This Section is devoted to a balance of the results obtained in
this paper. To make this balance crystal clear it is convenient to warn the
reader from mistaking our main purpose with that of many papers on the
subject of master equation and the foundation of stochastic Schr\"{o}dinger
equation. To establish this difference of perspective and purposes, we find
it convenient to make some comments on a representative group of papers.
These are those of Refs. \cite{PL77,D93,MG96,S96,DGS98,BKP98,JCW98}. 

We have to point out that it is possible, in principle, to derive
exact master equations to describe the time evolution of an open system
\cite{PL77}. However, in the specific case of the model of Caldeira and Leggett
\cite{cald}, which is the specific model studied in this letter, it is well known that the Markov approximation is responsible for the birth of unphysical
effects. This difficulty has been addressed by different authors with
different methods. For instance, Diosi \cite{D93} has recovered the Lindblad form \cite{lind} by adding two additional damping terms to the result of the Markov
approximation. Munro and Gardiner \cite{MG96} made the interesting observation that the unphysical effects might be the consequence of a transient process produced by the regression to equilibrium from the initial factorization  of the
system and bath's density operator. However, the reduced density matrix
fails being positive semidefinite in a short-time region which is probably
beyond experimental observation due to the coarse-grain time-scale approach
used in its derivation. This explicit admission of resting on a coarse-graining procedure is of significant importance to stress the main aim of
this letter, as we shall see in the remainder part of this Section.

We have to mention finally that the important problem of unravelling the master equation so as to build up an equivalent Schr\"{o}dinger
equation is now being currently extended with success to the case of
non-Markov master equations \cite{S96,DGS98,BKP98,JCW98}.

The main purpose of this paper is totally different from that of
these papers, but it is much more closely related to the conceptual issues
of \cite{penrose} and \cite{gpr90}. It is probably not so easy to appreciate these differences especially because the formal structures of the stochastic Schr\"{o}dinger equation of \cite{gpr90} might generate the false impression that these authors, and we with them, adopt a phenomenological approach rather than a more attractive derivation from ordinary quantum mechanics. 
        
First of all, we have to stress that our main thrust is on the
connection between the unification of quantum and classical
physics and the unification of mechanics and thermodynamics. The structure
of the stochastic Schr\"{o}dinger equation of Ghirardi, Pearle and Rimini \cite{gpr90} has been determined by the need of establishing a unified perspective embracing macroscopic classical physics. The spontaneous wave-function collapses are described by a correction to the ordinary Schr\"{o}dinger equation, compatible with the Lindblad structure \cite{lind}, but conceived as real correction to ordinary
quantum mechanics rather than as expression of the influence of an
environment. 
        We are convinced that to consider these corrections as
manifestations of the environment influence we should prove that the
Lindblad structure \cite{lind} can be derived from the quantum mechanical picture of a system interacting with an environment with no conflict with ordinary
quantum mechanics. This cannot be done because the Lindbland structure
\cite{lind} implies a rigorously exponential decay. When it is used, as we did in this paper, as a seed of stochasticity, it certainly kills the long-time
deviations from the exponential decay, in conflict with the well-known fact
that the exponential-like decay regime is only admitted in an intermediate
time region \cite{fonda}.

	The conviction that stochastic processes are compatible with
ordinary statistical mechanics is questionable from a conceptual perspective, 
even if it is especially attractive under the proviso of adopting the {\em for
all practical purposes} point of view. We know that the prototype of
stochastic processes is given by Brownian motion, and that the formal structure encompassing it within the usual statistical treatments is the ordinary Fokker-Plank equation. Therefore a convincing demonstration of the compatibility between ordinary quantum mechanics and stochastic processes would be given by a
derivation of the Fokker-Planck equation with no statistical or coarse-graining assumption whatsoever. The ordinary approach to the Fokker-Planck equation is not only affected by statistical assumptions such as averaging on the initial conditions: If the Markoffian assumption is not made, the resulting equation of motion cannot be identified with a {\em bona fide} Fokker-Planck equation \cite{f77}.
        
This does not rule out, of course, the possibility that a stochastic
process with infinite memory might exist. However, it can be shown \cite{GRW98}
that in this case this stochastic process turns out to be equivalent to
transport equations expressed in terms of fractional derivatives, thereby
departing from conventional statistical mechanics to which we would like to
limit the discussion of this final Section. 
        
The problem with advocating our point of view is that of the
experimental verification of the so called spontaneous fluctuations
\cite{BGMTV95}. Bonci {\em et al.} \cite{BGMTV95} have recently shown that the influence of the environment might not result, in principle, in a wave-function collapse, but only in a blurring of the wave function, even if from a statistical point of view wave-function collapses turn out to be indistinguishable from
decoherence processes with no collapse. A wave-function collapse is a single-system property which can be provoked by the environmental fluctuations only if these happen to be genuinely stochastic. Unfortunately, to experimentally prove this perspective it would be necessary to find cases where the effect of spontaneous fluctuations is statistically more significant than the environmental-induced fluctuations. We feel unconfortable in accepting a view where a wave-function collapse, requiring the action of a genuine stochastic process, is
actually caused by our ignorance of the microscopic details, or more in
general by the limitations of the human observer. However, to change a
philosophical debate into a scientific issue it would be necessary to single
out experimental effects, and in this sense the results of \cite{BGMTV95} are not encouraging. In fact, the toy model adopted to study quantum jumps show that
the rate of the spontaneous wave-function collapses is much lower than that
of the environment-induced decoherence, thereby making them virtually
invisible to statistical observation: A nice price to pay to leave quantum
statistical mechanics essentially unchanged. 
        
We can show, however, that in spite of the discouraging  conditions
concerning the experimental settlement of this problem, the adoption of our
perspective can lead to the definition of new interesting theoretical problems, concerning the triggering action of stochastic processes, regardless the philosophical view of the investigator. A nice example of this kind is provided by Zurek and Paz \cite{zupaz}. These authors address indeed the problem of the foundation of the second principle of thermodynamics by means of the same theoretical arguments as those adopted by Zurek to address the intriguing problem of the
wave-function collapse \cite{Z91}. Although we do not share Zurek's view, since
we think it to be impossible to derive white noise from the contraction over
the environmental degrees of freedom, we find attractive the ensuing picture
of how the influence of this randomness, which in our perspective would be a
spontaneous fluctuations, is  enhanced by the chaotic properties of the
system under study.
        
It has to be pointed out that much of the confidence of the
advocates of the foundation of thermodynamics on the basis of ordinary
physics rests on the identification of the correlation time $\tau_M$ with the
predictability time $\tau_S$. Both times are defined in the paper of Ref. \cite{GGTV96}, which has been devoted to shed light on this intriguing issue, and where it was proved that the conventional view is reinforced by the key role of the following inequality
\be
\tau_M \ll \tau_S \ll \frac{1}{\eta}. \label{nw}     
\ee
This is Eq. (24) of Ref. \cite{GGTV96} with the oscillator friction denoted by $\eta$ to fit the notation of this letter. If this inequality is fulfilled, then the resulting macroscopic stochastic
dynamics become indistinguishable from those predicted by ordinary physics
supplemented by the arbitrary assumption of identifying $\tau_M$ with the
predictability time $\tau_S$. Actually $\tau_M$ is the lifetime of a correlation function, and this is not necessarily identical to the time at which the
system under study stops being predictable. An illuminating example is given
by the superposition of many harmonic oscillations with slight different
frequencies. The resulting process might be characterized by a
relaxation-like behavior, but it would mistifying to identify the resulting
decorrelation time with the beginning of an unpredictable regime \cite{GGTV96}. In classical physics this unpredictability time is proved to be inversely
proportional to the Lyapunov coefficient. Apparently there might be a
conflict between this interpretation and the fact that other investigators
might identify this time with the inverse of the correlation time $1/\tau_M$.
This apparent conflict is settled by adopting the perspective of Zurek and
Paz \cite{zupaz}: The key action of the environmental seed of irreversibility makes it possible to identify the two times, since in the case studied by Zurek
and Paz $\tau_M$ is slightly shorter than $\tau_S$. This is in a perfect accordance with the point of view that we are trying to illustrate here, even if are
inclined to identify the seed of irreversibility with spontaneous
fluctuations either of the form of those of Ghirardi, Pearle and Rimini, or
of the Mensky constrained path integral studied in this letter.
        
The main aim of this paper has been that of assessing whether or not
the adoption of the Mensky path integral with constraint, as an expression
of correction to the ordinary quantum dynamics of the bath oscillators,
makes it possible to realize the basic condition (\ref{nw}). This condition, fully expressed in terms of the notations used in this paper, would read
\be
\eta < \Gamma < \Omega.  \label{nnw} 
\ee
In fact, $1/\Omega$ plays the same role as ordinary correlation time, and the
parameter $\Gamma$ can be adopted to denote the rate of the Mensky process on a
single oscillator. Note that the Mensky measurement process is the
R-process of the perspective established by Penrose \cite{penrose}, and consequently the time $\tau_S$ must be identified with $1/\Gamma$, since the measurement process is by definition a genuinely stochastic process and the calculations of Section III have shown that a single bath oscillator is made unpredictable
in time with the rate $\Gamma$.

Note that establishing this inequality, as simple as it is, cannot
be easily done without the extended calculations of Section III, and
especially without the complex process of repeated trials yielding Eq. (\ref{kappa}). Following \cite{tvg95}, we adopt for the dissipation parameter of the oscillator, $\eta$, the value $10^9$ sec$^{-1}$. This plausible choice and the inequality (\ref{nnw}) set a minimum value for the parameter $\Gamma$, which turns out too large to be compatible with the Taylor series expansion adopted in our theoretical treatment. Furthermore, if we fix $\Gamma = 10^{10}$ sec$^{-1}$, $m_B = 10^{-24}$ gr, $\Omega = 10^{12}$ sec$^{-1}$, we find that the real part of $k$ is equal to the value $10^{25}$ sec$^{-1}$ cm$^{-2}$. This means that the corrections made are unacceptable from the physical as well as from the mathematical point of view. In other words, the results are not selfconsistent and imply corrections to ordinary
physics too large to be seriously taken into account. The spontaneous
fluctuations of Ghirardi, Pearle and Rimini \cite{gpr90} lead to much smaller
corrections, and consequently might be regarded as being in principle an
acceptable corrections. However, as established in \cite{tvg95}, in spite of the enhancement produced by the interplay with the ordinary
fluctuation-dissipation process, these fluctuations fail producing the
objective wave-function collapse in a reasonably short time, thereby sharing
the weakness of the Mensky constrained path integral studied in this letter.
        
According to the perspective here adopted, that the Markov
assumption corresponds to tacitly correcting quantum mechanics, this
conclusion seems to be equivalent to establishing that the Markov
approximation made in literature are unacceptable. However, suggesting that
all master equations based on this assumption, widely used in literature,
cannot be trusted, would be a dramatic conclusion. We are convinced that the
results of this letter lead to a rather different conclusion that will be
illustrated at the end as the main result of this paper. Before reaching
this conclusion, we want to caution the reader from mistaking this
conclusion for the discovery that the Markov assumption can break the
condition that the density matrix is positive definite \cite{D93,MG96}. We
cannot rule out completely the possibility that the perspective of this
paper can be used to shed light on some issues raised, for instance, by
Munro and Gardiner \cite{MG96}\footnote{We refer here to some key equations of Ref. \cite{MG96}. It seems to us that the short-time expansion of the density matrix terms of Eqs. (3.3) and (3.5) with the joint use of (3.11) might lead to a breakdown of the property of the density matrix to be positive definite in a time region more extended than the correlation time of Eq. (2.20). This seems to support our view that the correlation time $\tau_M$, identified by us with the time of Eq. (2.20), cannot be regarded as the true coarse-graining time. Averages should be made on times more extended, as indicated by Eq. (\ref{nnw}) of this letter.}. However, our thrust focuses on the different issue widely illustrated in this Section.
        
We are convinced that the results of this letter leads to a
conclusion different from stating the breakdown of the master equations. The
master equations, supplemented in case with the improvements established by
\cite{D93} and \cite{MG96}, might lead to correct result, in spite of the fact that Eq. (\ref{nnw}) seems to be unphysical. It has to be remarked that the approach to statistical mechanics here adopted is not genuinously dynamic: the bath of oscillators is linear and it is arbitrarily forced to stay in a canonical
equilibrium condition. For this reason the conventional assumption that the
bath oscillators are in a state of canonical equilibrium has to be replaced
by dynamical ingredients (see, for instance, the discussion made in recent
literature \cite{bia,hoo}). This means that each bath oscillator must be assigned its own bath, and this bath must be non-linear so as to result, in the
classical limit, in deterministic chaos. This is expected to lead to the
fulfillment of Eq. (\ref{nnw}) without internal inconsistencies, thereby explaining why, after all, the linear baths of literature can be profitably used with
no conflict with reality.
        
This is a plausible conjecture that has to be proved by a direct use
of a nonlinear bath. There are already outstanding examples of dynamic
approaches to quantum statistical mechanics, based on the so called quantum
chaos \cite{s94,fi97}. We note that within this context it has already been noticed \cite{zupaz} that a chaotic dynamic system leads to a significant increase of the Gibbs entropy in times of the order of the inverse of the Lyapunov
coefficients, provided that a seed of genuine irreversibility is
introduced as an effect of environmental fluctuations. As earlier said, this
fits very well the perspective established by the present paper. The
dependence on these environmental fluctuations is through the logarithm of
their intensity, thereby implying no significant change to ordinary quantum
mechanics. If we interpret these environmental fluctuations as weak
corrections to ordinary quantum mechanics, of the same type as that here
studied, we might reach the satisfactory conclusion that a non arbitrary
Markov assumption can be realized with extremely weak corrections to
ordinary quantum mechanics. The results of this paper therefore must be
interpreted as an incentive to study directly, from now on, the influence that white fluctuations have on dynamic systems which would be strongly chaotic in the classical limit.


\begin{references}
\bibitem{penrose} R. Penrose, Shadows of the Mind - A Search for the Missing
Science of Consciousness, Oxford University Press, Oxford, 1994.

\bibitem{bl85}  A. Barchielli and G. Lupieri, J. Math. Phys. 26 (1985) 2222.

\bibitem{ba86}  A. Barchielli, Phys. Rev. A 34 (1986) 1642.

\bibitem{be89}  V.P. Belavkin, Phys. Lett. A 140 (1989) 355; J. Phys. A:
Math. and Gen. 22 (1989) L1109.

\bibitem{fh}  R.P. Feynman and A.R. Hibbs, Quantum Mechanics and Path
Integrals, McGraw-Hill, NY, 1965.

\bibitem{mensky}  M. Mensky, Phys. Lett. A196 (1994) 159.

\bibitem{pot}  C. Presilla, R. Onofrio and U.Tambini, Ann. Phys. 248 (1996)
95.

\bibitem{pot2}  C. Presilla, R. Onofrio and M. Patriarca, J. Phys. A30
(1997) 7385.

\bibitem{fv}  R.P. Feynman and F.L. Vernon, Ann, Phys. 24 (1963) 118.

\bibitem{tvg95}  L. Tessieri, D. Vitali, P. Grigolini, Phys. Rev. A 51
(1995) 4404.

\bibitem{a76}  S.A. Adelman, J. Chem. Phys. 64 (1976) 124.

\bibitem{m65}  H. Mori, Prog. Theor. Phys. 33 (1965) 423.

\bibitem{m65b}  H. Mori, Prog. Theor. Phys. 34 (1965) 399.

\bibitem{f77}  R.F. Fox, J. Math. Phys. 18 (1977) 2331.

\bibitem{ks95}  H. Kleinert, S.V. Shabanov, Phys. Lett. A 200 (1995) 224.

\bibitem{ds97}  L. Di\'{o}si and W.T. Strunz, quant-ph/9706050.

\bibitem{vitali}  D. Vitali and P. Grigolini, Phys. Rev. A39 (1989) 1486.

\bibitem{lee}  M.H. Lee, Phys. Rev. Lett. 51 (1983) 1227.

\bibitem{fonda}  L. Fonda, G.C. Ghirardi and A. Rimini, Rep. Prog. Phys. 41
(1978) 587.

\bibitem{mg98}  A. Mazza, P. Grigolini, Phys. Lett. A 238 (1998) 169.

\bibitem{cald}  A.O. Caldeira and A.J. Leggett, Physica A 121 (1983) 587.

\bibitem{m87}  M. B. Mensky, Phys. Lett. A 231 (1997).

\bibitem{gpr90}  G.C. Ghirardi, P. Pearle, and A. Rimini, Phys. Rev. A 42
(1990) 78.

\bibitem{cald2}  A.O. Caldeira and A.J. Leggett, Ann. of Phys. 149 (1983)
374.

\bibitem{west}  K. Lindenberg, B.J. West, Phys. Rev. A30 (1984) 568.

\bibitem{hpz}  B.L. Hu, J.P. Paz and Y. Zhang, Phys. Rev. D 45 (1992) 2843.

\bibitem{grab}  H. Grabert, P. Schramm and G.-L. Ingold, Phys. Rep. 168
(1988) 115.

\bibitem{pat}  M. Patriarca, Nuovo Cimento B 111 (1996) 61.

\bibitem{g85}  P. Grigolini, Adv. Chem. Phys. 62 (1985) 1.

\bibitem{ja}  E.T. Jaynes, Phys. Rev. 106 (1957) 620; Phys. Rev. 108 (1957)
171.

\bibitem{PL77} R.R. Puri and S.V. Lawande, Phys. Lett. A 62 (1977) 143.

\bibitem{D93} L. Diosi, Physica A 199 (1993) 517.

\bibitem{MG96} W.J. Munro and C.W. Gardiner, Phys. Rev A 53 (1996) 2633.

\bibitem{S96} W.T. Strunz, Phys. Lett. A 224 (1996) 25.

\bibitem{DGS98} L. Diosi, N. Gisin, and W.T. Strunz, Phys. Rev. A 58 (1998) 1699.

\bibitem{BKP98} H.P. Breuer, B. Kappler, and F. Petruccione, quant-ph/9806026.

\bibitem{JCW98} M.W. Jack, M.J. Collett, and D.F. Walls, quant-ph/9807028.

\bibitem{lind} G. Lindblad, Comm. Math. Phys. 48 (1976) 119.

\bibitem{GRW98} P. Grigolini, A. Rocco, B.J. West, cond-mat/9809075.

\bibitem{BGMTV95} L. Bonci, P. Grigolini, G. Morabito, L. Tessieri, and D. Vitali, Phys. Lett. A 209 (1995) 129.

\bibitem{zupaz} W.H. Zurek and J.P. Paz, Phys. Rev. Lett. 72 (1994) 2508.

\bibitem{Z91} W.H. Zurek, Physics Today 44 (10) (1984) 87.

\bibitem{GGTV96} V. Giovannetti, P. Grigolini, G. Tesi and D. Vitali, Phys. Lett. A 224 (1996) 31.

\bibitem{bia}  M. Bianucci, R. Mannella, B.J. West, P. Grigolini, Phys. Rev. E 51, (1995) 3002.

\bibitem{hoo} Wm.G. Hoover, Computational Statistical Mechanics, Elsevier, Amsterdam, 1991.

\bibitem{s94}  M. Srednicki, Phys. Rev. E 50 (1994) 888.

\bibitem{fi97}  V.V. Flambaun and F.M. Izrailev, Phys. Rev. E 56 (1997) 5144.


\end{references}
\end{document}